\documentclass[journal,twoside,final]{IEEEtran}

\usepackage{array}

\usepackage{amsmath}

\usepackage{subfigure}
\usepackage{graphicx}
\usepackage{amsmath}
\usepackage{amssymb}
\usepackage{mathrsfs}
\usepackage{citesort}
\usepackage{longtable}
\usepackage{url}
\usepackage{float}
\usepackage{acronym}
\usepackage{paralist}

\acrodef{SMlow}[SM]{spatial modulation}
\acrodef{SMcap}[SM]{Spatial modulation}
\acrodef{FBElow}[FBE]{fractional bit encoding}
\acrodef{FBE--SM}{fractional bit encoded spatial modulation}
\acrodef{MIMOlow}[MIMO]{multiple--input multiple--output}
\acrodef{MIMOcap}[MIMO]{Multiple--input multiple--output}
\acrodef{ICIlow}[ICI]{inter--channel interference}
\acrodef{ml}[ML]{maximum likelihood}
\acrodef{sulb}[SULB]{single-user-lower-bound}
\acrodef{mimolow}[MIMO]{multiple--input multiple--output}
\acrodef{mimocap}[MIMO]{Multiple--input multiple--output}
\acrodef{ber}[BER]{bit-error-ratio}
\acrodef{SER}{symbol error ratio}
\acrodef{AF}{amplify and forward}
\acrodef{DF}{decode and forward}
\acrodef{IAI}{inter--antenna interference}
\acrodef{aber}[ABER]{average bit error ratio}
\acrodef{BER}{bit error ratio}
\acrodef{aser}[ASER]{average symbol error ratio}
\acrodef{ssk}[SSK]{space shift keying}
\acrodef{snr}[SNR]{signal-to-noise-ratio}
\acrodef{siso}[SISO]{single-input single-output}
\acrodef{simo}[SIMO]{single-input multiple-output}
\acrodef{sinr}[SINR]{signal-to-interference-plus-noise-ratio}
\acrodef{psk}[PSK]{phase shift keying}
\acrodef{QAM}{quadrature amplitude modulation}
\acrodef{MA}{multiple active}
\acrodef{pep}[PEP]{pairwise error probability}
\acrodef{iid}[i.i.d.]{identical and independently distributed}
\acrodef{csi}[CSI]{channel state information}
\acrodef{STBC}{space-time block codes}
\acrodef{PAM}{pulse amplitude modulation}
\acrodef{SIC}{successive interference cancellation}
\acrodef{gsm}[GSM]{generalised spatial modulation}
\acrodef{IGCH}{information guided channel hopping}
\acrodef{RF}{radio frequency}
\acrodef{FDMA}{frequency division multiple access}
\acrodef{TDMA}{time division multiple access}
\acrodef{OFDMA}{orthogonal frequency division multiple access}
\acrodef{AWGN}{additive white Gaussian noise}
\acrodef{AF}{amplify and forward}
\acrodef{DF}{decode and forward}
\acrodef{Dh-SM}{dual-hop spatial modulation}
\acrodef{DSM}{distributed spatial modulation}
\acrodef{ARQ}{automatic repeat request}
\acrodef{MRC}{maximum ratio combining}
\acrodef{MGF}{moment generating function}
\acrodef{DDF}{dynamic decode and forward}
\acrodef{LoS}{line of sight}
\acrodef{NLoS}{non--line of sight}
\acrodef{CDF}{cumulative density function}
\acrodef{RRC}{root raised cosine}
\acrodef{FIR}{finite impulse response}
\acrodef{D-BLAST}{Diagonal Bell Laboratories Layered Space-Time Architecture}
\acrodef{V-BLAST}{Vertical Bell Laboratories Layered Space-Time Architecture}
\acrodef{MPPI}{Moore-Penrose pseudo-inverse}
\acrodef{SVD}{sigular value decomposition}
\acrodef{MMSE}{minimum mean-squared error}
\acrodef{ZF}{zero-forcing}
\acrodef{SD}{sphere decoder}
\acrodef{TOSD-SM}{time-orthogonal signal design assisted spatial modulation}
\acrodef{TOSD-SSK}{time-orthogonal signal design assisted space shift keying}
\acrodef{OFDM}{orthogonal frequency division multiplexing}
\acrodef{GSMC}{Global System for Mobile Communications }
\acrodef{ITU}{International Telecommunications Union}
\acrodef{LAN}{Local Area Network}
\acrodef{CDMA}{code division multiple access}
\acrodef{LTE-A}{Long-Term Evolution Advanced}
\acrodef{TAS}{transmit-antenna selection}
\acrodef{SDM}{spatial division multiplexing}
\acrodef{BPSK}{binary phase shift keying}
\acrodef{QPSK}{quadrature phase shift keying}
\acrodef{DSP--Tx}{digital signal processing at the transmitter}
\acrodef{DSP--Rx}{digital signal processing at the receiver}
\acrodef{PXIe--Rx}{NI--PXIe chassis at the receiver}
\acrodef{PXIe--Tx}{NI--PXIe chassis at the transmitter}
\acrodef{FO}{frequency offset}
\acrodef{LS}{least squares}
\acrodef{SMX}{spatial multiplexing}
\acrodef{IFI}{inter--frame interference}
\acrodef{ISI}{inter--symbol interference}
\acrodef{ni}[NI]{National Instruments}
\acrodef{pim}[PI]{power imbalance}
\acrodef{bs}[BS]{base station}

\newcommand{\bs}{\ac{bs}}
\newcommand{\pim}{\ac{pim}}
\newcommand{\nis}{\ac{ni}}
\newcommand{\isi}{\ac{ISI}}
\newcommand{\ifi}{\ac{IFI}}
\newcommand{\smx}{\ac{SMX}}
\newcommand{\ls}{\ac{LS}}
\newcommand{\fo}{\ac{FO}}
\newcommand{\pxietx}{\ac{PXIe--Tx}}
\newcommand{\pxierx}{\ac{PXIe--Rx}}
\newcommand{\dsprx}{\ac{DSP--Rx}}
\newcommand{\dsptx}{\ac{DSP--Tx}}

\newcommand{\bpsk}{\ac{BPSK}}
\newcommand{\fbesm}{\ac{FBE--SM}}

\newcommand{\sd}{\ac{SD}}

\newcommand{\fir}{\ac{FIR}}
\newcommand{\rrc}{\ac{RRC}}
\newcommand{\cdf}{\ac{CDF}}
\newcommand{\los}{\ac{LoS}}
\newcommand{\nlos}{\ac{NLoS}}

\newcommand{\ie}{{\it i.e.}}

\newcommand{\aber}{\ac{aber}}

\newcommand{\snr}{\ac{snr}}
\newcommand{\rf}{\ac{RF}}

\newcommand{\awgn}{\ac{AWGN}}

\newcommand{\qam}{\ac{QAM}}
\newcommand{\ssk}{\ac{ssk}}

\newcommand{\pep}{\ac{pep}}
\newcommand{\gsm}{\ac{gsm}}

\newcommand{\ml}{\ac{ml}}

\newcommand{\simo}{\acs{simo}}
\newcommand{\mimo}{\ac{mimolow}}

\newcommand{\sm}{\ac{SMlow}}
\newcommand{\ici}{\ac{ICIlow}}

\usepackage{stfloats}

\usepackage{flushend}

\acresetall 

\begin{document}

\title{\huge Practical Implementation of Spatial Modulation}
\author{
	N.~Serafimovski,
	A.~Younis,
        R.~Mesleh, 
	P.~Chambers, 
	M.~Di~Renzo, 
	C.-X.~Wang,
	P.~M.~Grant, \\
	M.~A.~Beach,
        and H.~Haas 
\thanks{%
Copyright (c) 2013 IEEE. Personal use of this material is permitted. However, permission to use this material for any other purposes must be obtained from the IEEE by sending a request to pubs-permissions@ieee.org. 
} 
\thanks{%
The associate editor coordinating the review of this paper and approving it for publication was Prof. Yong Liang Guan. Manuscript received October 2, 2012; revised February 18, 2013 and April 24, 2013.
}
\thanks{%
 N.~Serafimovski, A.~Younis, P.~M.~Grant and H.~Haas are with The
University of Edinburgh, Edinburgh, EH9 3JL, UK, (e--mail: \{n.serafimovski,
a.younis, p.grant, h.haas\}@ed.ac.uk.).
}
\thanks{%
 M.~Di~Renzo is with the Laboratoire des Signaux et Syst\`emes, Unit\'e Mixte de Recherche 8506, Centre National de la Recherche Scientifique--\'Ecole Sup\'erieure d'\'Electricit\'e--Universit\'e Paris--Sud XI, 91192 Gif--sur--Yvette Cedex, France, (e--mail: marco.direnzo@lss.supelec.fr).
}
\thanks{%
 R.~Mesleh is with the Electrical Engineering Department and SNCS research center, University of Tabuk, P.O.Box: 71491 Tabuk, Saudi Arabia, (e--mail: rmesleh.sncs@ut.edu.sa).
}
\thanks{%
 P.~Chambers and C.-X. Wang are with Heriot-Watt University, Edinburgh, EH14 4AS, UK. (e--mail: 
\{P.Chambers, Cheng-Xiang.Wang\}@hw.ac.uk).
}
\thanks{%
 M. A. Beach is with The University of Bristol, Bristol, BS8 1UB, UK. (e--mail: 
M.A.Beach@bristol.ac.uk)
}
\thanks{%
 Digital Object Identifier 00.0000/TVT.0000.00.000000 
}
}

\IEEEpubid{0000--0000/00\$00.00˜\copyright˜0000 IEEE}

 \markboth{IEEE TRANSACTIONS ON VEHICULAR TECHNOLOGY, ACCEPTED FOR PUBLICATION}{Serafimovski \MakeLowercase{\textit{et al.}}: Practical Implementation of Spatial Modulation}

\maketitle

\begin{abstract}
In this work we seek to characterise the performance of \sm\ and \smx\ with an experimental testbed. 
Two National Instruments (NI)-PXIe devices are used for the system testing, one for the transmitter and one for the receiver. 
The digital signal processing that formats the information data in preparation for transmission is described along with the digital signal processing that recovers the information data. In addition, the hardware limitations of the system are also analysed. 
The \aber\ of the system is validated through both theoretical analysis and simulation results for \sm\ and \smx\ under \los\ channel conditions.
\end{abstract}

\acresetall 

\acresetall  

\section{Introduction}
\label{sec:testbed_intro}

\begin{figure} 
 \centering
 \includegraphics[width=8.3cm]{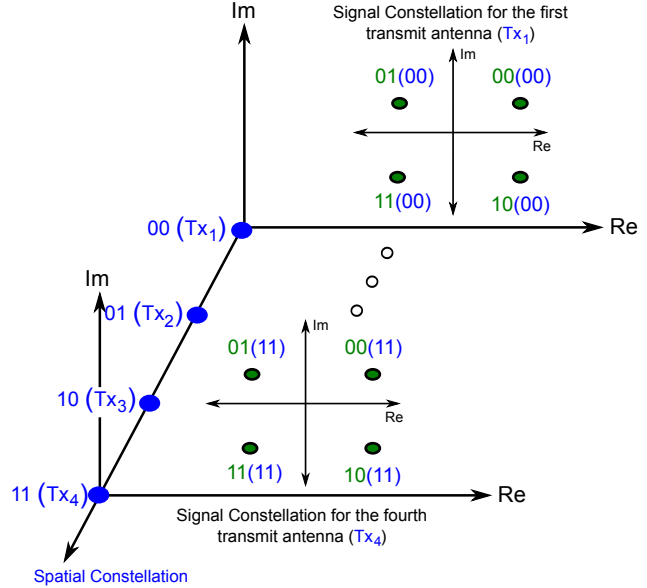}
 \caption{The unique three dimensional constellation diagram for \sm. 
The lower two bits, in the four bit word define the spatial--constellation point which identifies the active antenna. These are shown in parentheses. The remaining two bits determine the signal--constellation point that is to be transmitted.}
 \label{fig:SM_example}
   \vspace{-0.5cm}
\end{figure}

 Multiple--input multiple--output (MIMO)\acused{mimolow} systems offer a significant increase in spectral efficiency in comparison to single antenna systems~\cite{t9902,mslgh0901}. An example is \sm, which increases the spectral efficiency of single antenna systems while avoiding \ici~\cite{mhly0501}. This is attained as shown in Fig.~\ref{fig:SM_example}, through the adoption of a new modulation and coding scheme, which foresees: i)~the activation, at each time instance, of a single antenna that transmits a given data symbol (\emph{constellation symbol}), and ii)~the exploitation of the spatial position (index) of the active antenna as an additional dimension for data transmission (\emph{spatial symbol})~\cite{mhsay0801}. Both the \emph{constellation symbol} and the \emph{spatial symbol} depend on the incoming data bits. An overall increase by the base--two logarithm of the number of transmit--antennas of the spectral efficiency is achieved. This limits the number of transmit antennas to be a power of two unless~\fbesm~\cite{srsmh1001}, or~\gsm~\cite{ysmh1001} are used. In particular, in~\cite{ysmh1001} it is shown that the number of spatial symbols does not need to be equal to the number of transmit antennas. For example, if \gsm\ is used, the number of spatial symbols is equal to the number of unique channel signatures between the transmitter and receiver, where the unique channel signatures can be obtained by activating various combinations of the available transmit antennas. In this work, however, these unique channel signatures are assumed to be due to the activation of individual transmit antennas.

\begin{figure*}[ht]
    \centering
      \includegraphics[scale=0.45]{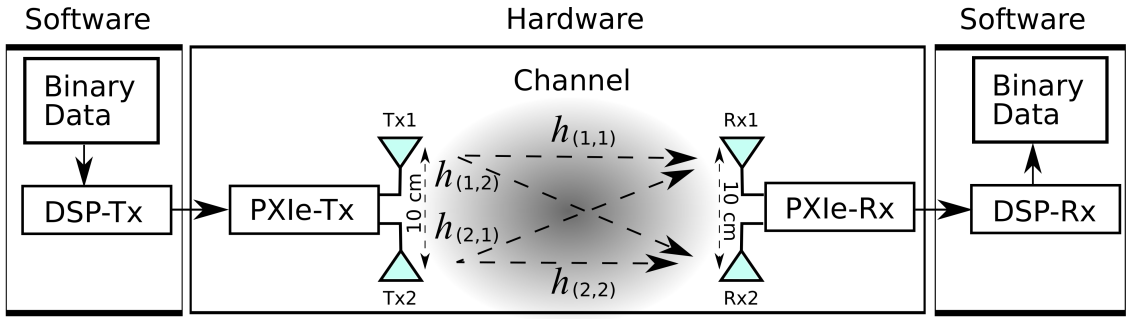} 
      \caption{ Block sequence of the main steps in the experiment, from the generation of the binary data to its recovery. }
    \label{fig:Block1}
\end{figure*}

\IEEEpubidadjcol

Activating only one antenna at a time means that only one RF chain is needed, which significantly reduces the hardware complexity of the system~\cite{jgsc0901}. Moreover, the most energy consuming part of a base station are the power amplifiers and the \rf\ chains associated with each transmitter \cite{agdgsoisg1101}, where the power requirements of a base station are shown to increase linearly with the number of \rf\ chains added  \cite{ddgfahwsr1201}. However, as only one RF chain is needed, SM offers a reduction in the energy consumption which scales linearly with the number of transmit antennas~\cite{ssrhg1201,ssrh1201}. Furthermore, the computational complexity of SM--ML is equal to the complexity of single--input multiple--output (SIMO) systems~\cite{jgs0801}, \emph{i.e.} the complexity of SM--ML depends only on the spectral efficiency and the number of receive antennas, and does not depend on the number of transmit antennas. Moreover, in~\cite{ysrmh01,ymhg1001,yrmh1101}, the complexity of \sm\ is further reduced by using the~\sd.

Several papers that seeks to understand and hence improve the performance of \sm\ in various scenarios are available in literature. In~\cite{mrhg1001,rh1004}, the \aber\ performance of \sm\ is improved by introducing trellis coding on the transmitting antennas. The optimal  detector is derived with and without channel state information at the receiver in~\cite{jgs0801,hjls0901,dh1001}. The \aber\ performance is given when considering channel estimation errors in~\cite{im1201,bupp1201,rlgh1201}. The optimal power allocation for the case of two transmit antennas and one receive antenna system is given in closed form in~\cite{rh1006}, and the \aber\ performance of \sm\ in correlated fading channels is considered in~\cite{hms0901}. In~\cite{bapp1101,dh1301a} spectral efficiency and diversity gains are obtained by combining \sm\ with space-time block codes~(STBC--SM). Applying \sm\ to relaying systems is also shown to result in significant \snr\ gains when compared to non-cooperative decode and forward techniques~\cite{ssrh1101}. 
 In~\cite{ssrh1309} the overall power performance of a \bs\ employing \sm\ is studied.
 More recently, a comprehensive analytical framework to compute the \aber\ of \sm\ over generalized fading channels is introduced in~\cite{dh1201a}. Moreover, in~\cite{ytrwbhg1301} for the first time the performance of \sm\ is analysed using real--world channel measurements. The latest research achievements and an outline of some relevant open research issues for \sm\ are reviewed in~\cite{rhg1101}. All research thus far is strictly theoretical.
 
 In this paper, the \aber\ performance of \sm\ is analysed in a practical testbed and compared with that for \smx. In particular, the \nis--PXIe--1075 chassis are used at the transmitter and receiver. The design of the testbed hardware and the software used are explained in detail along with the transmission chain. 
 The effects of the entire transmission chain on the system performance are examined. The basic elements of the transmission link are the transmit \rf\ chain, the wireless channel, and the receive \rf\ chain. In addition to the effects of the wireless channel on the phase and amplitude of the signal, the impact on the system performance of the power imbalances (PIs)\acused{pim} in the transmitter and receiver \rf\ chains is discussed.
 Furthermore, an analytical upper bound for the \aber\ performance of \sm\ over \nlos\ channels with \pim\ is derived, and compared to the experimental and computer simulation results. The experimental results validate the analytical bound as well as the attained computer simulations. Finally the performance of \sm\ is compared with the theoretical and experimental results of \smx.

 This paper is organised as follows.
The system set-up, equipment and digital signal processing are presented in Section~\ref{sec:testbed_sys_model}. 
The equipment constraints are then considered in Section~\ref{sec:testbed_constraints}, while the analytical modelling is discussed in Section~\ref{sec:testbed_ana}. In addition, the computational complexity of the \sm\ decoder algorithm is presented in Section~\ref{sec:complexity}. The performance of \sm\ is then characterised in the experimental and simulation environments in Section~\ref{sec:testbed_results}, where it is compared with the theoretical and experimental results of a \smx\ system. Lastly, the paper is summarised in Section~\ref{sec:testbed_conclusion}.

\section{Testbed Set--up and System Model} \label{sec:testbed_sys_model}

The testbed set--up and transmission chain can be separated into software and hardware parts, as shown in Fig.~\ref{fig:Block1}. The hardware consists of the \pxietx\ and the \pxierx. The software consists of the \dsptx\ and the \dsprx.

 The binary data to be broadcast is first processed by \dsptx, before being transmitted through the fading channel by the \pxietx. 
 The channel coefficient on the link between transmit antenna $n_t$, and receive antenna $r$, is denoted by $h_{\left(r,n_t\right)}$. Note that the number of antennas at the transmitter and the receiver are denoted by $N_t$ and $N_r$, respectively. At the receiver, the \pxierx\ records the \rf\ signal and passes it through to the \dsprx\ for processing, where the original data stream is recovered.
 
 \subsection{Testbed Hardware}
 
\begin{figure}[t]
\centering
\begin{minipage}[t]{0.49\linewidth}
  \subfigure[PXIe--Tx]{
    \label{fig:PXIe--Tx}
    \includegraphics[width=\linewidth]{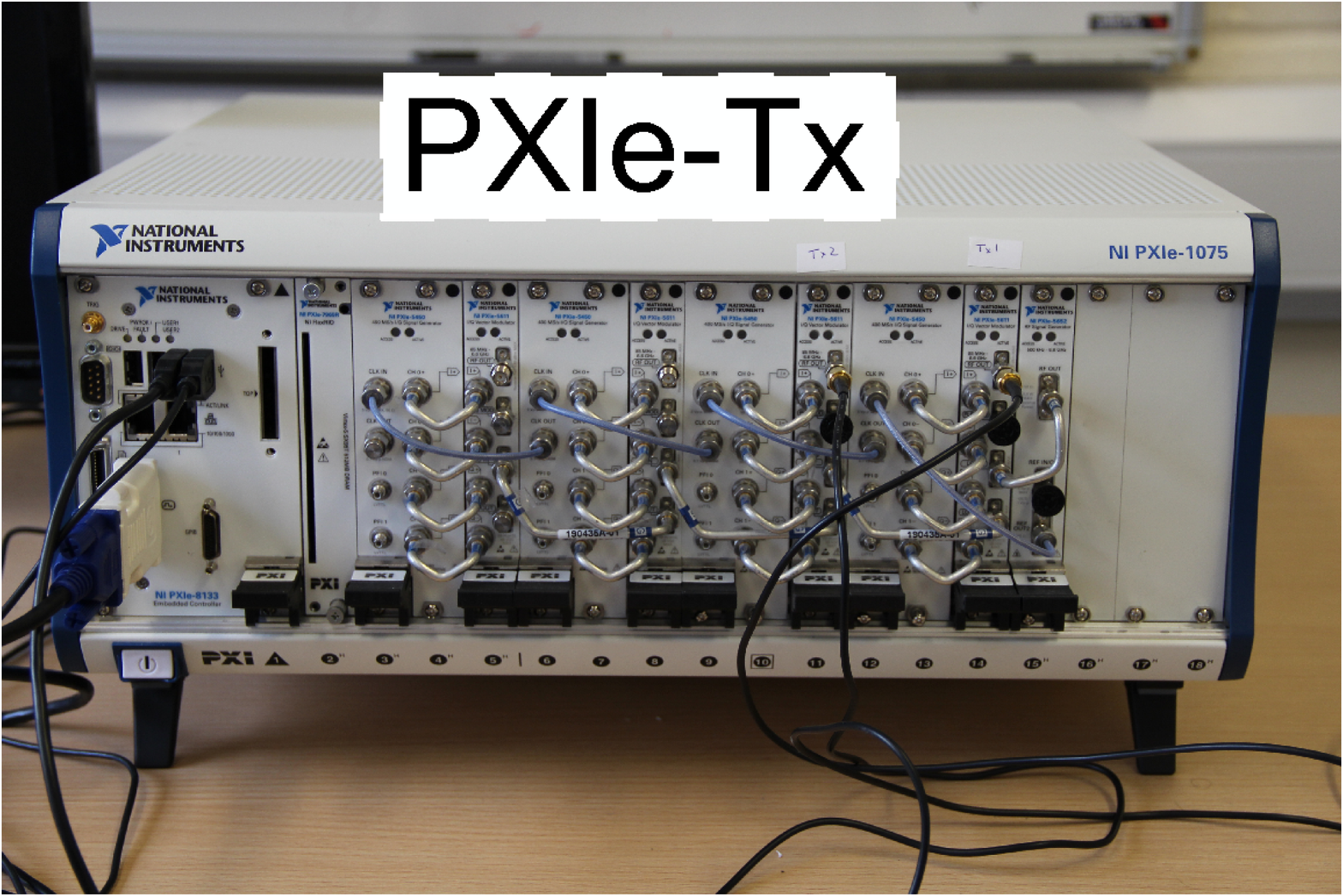} 
  } \\
\end{minipage}
\hfill
\begin{minipage}[t]{0.49\linewidth}
  \subfigure[PXIe--Rx]{
    \label{fig:PXIe--Rx}
    \includegraphics[width=\linewidth]{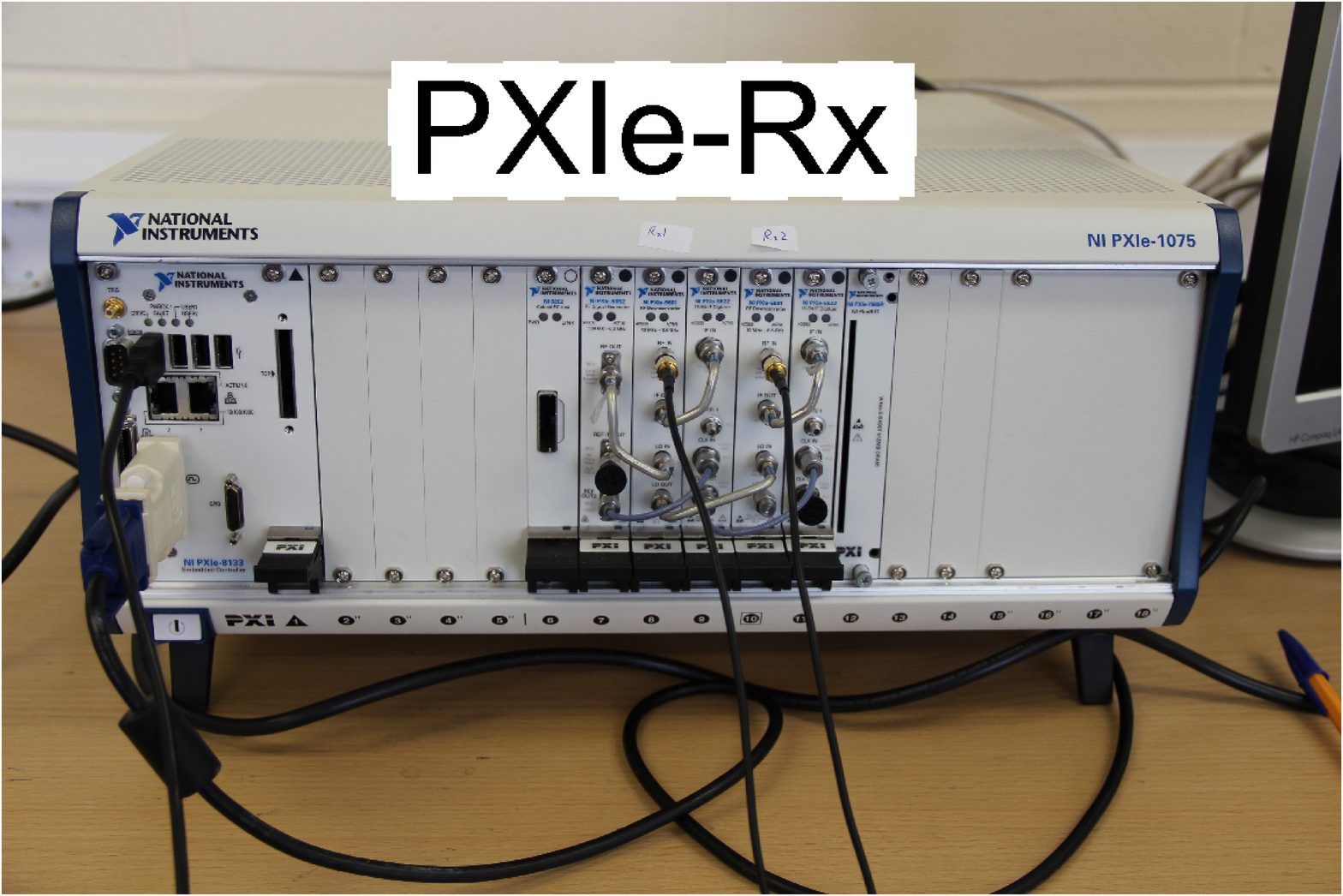}
  }
\end{minipage}
  \caption{NI--PXIe--1075 chassis with the relevant on--board modules at the transmitter (PXIe--Tx), and at the receiver (PXIe--Rx). }
  \label{fig:PXIe}
\end{figure}

 The NI--PXIe--1075 chassis are equipped with a $1.8$~GHz Intel--i$7$ processor with $4$~GB RAM and are shown in Fig.~\ref{fig:PXIe}. 
 The system has two transmit antennas and two receive antennas. Each antenna at the transmitter and receiver contains two quarter--wave dipoles, and one half--wave dipole placed in the middle. All three dipoles are vertically polarised. In addition, each antenna has a peak gain of $7$~dBi in the azimuth plane, with an omnidirectional radiation pattern.

\subsubsection{Transmitter hardware (PXIe--Tx)} \label{subsubsec:testbed_physical_Tx}

The following NI--PXIe modules are used at the transmitter,

\begin{itemize}
\item NI-PXIe-5450 16-Bit I/Q Signal Generator (SG--16bit),
\item NI-PXIe-5652 \rf\ Signal Generator with a $500$~kHz to $6.6$~GHz frequency range (SG--RF),
\item NI-PXIe-5611 intermediate frequency (IF) to carrier \rf\ up--converter (up--converter).
\end{itemize}

The \pxietx\ has an operational frequency range of $85$~MHz to $6.6$~GHz and can facilitate a bandwidth of $100$~MHz at a maximum transmission power of $5$~dBm.

At the transmitter, the SG--16bit  performs a linear mapping of the signed $16$-bit range to the output power and polarisation, \ie, the peak voltage amplitude is assigned to any value in the transmission vector equal to $2^{15}$ with a linear scale of the voltage amplitude down to zero. The output from SG--16bit  then goes to SG--RF, which is connected to the up--converter. The up--converter outputs the analogue waveform corresponding to the data resulting from \dsptx\ at a carrier frequency of $2.3$~GHz. This completes a single \rf\ chain. The transmission of the \rf\ signal by the up--converters is synchronised by using a $10$~MHz reference signal.

\subsubsection{Receiver hardware (PXIe--Rx)} \label{subsubsec:testbed_physical_Rx}

The following NI--PXIe modules are used at the receiver,

\begin{itemize}
\item NI-PXIe-5652 an on-board reference clock (SG--RF),
\item NI-PXIe-5622 16-Bit Digitiser which records data samples in an I16 format (16-Bit Digitiser),
\item NI-PXIe-5601 \rf\ down--converter (down--converter).
\end{itemize}

The \pxierx\ can operate in a frequency range of $10$~MHz to $6.6$~GHz and can facilitate an operational bandwidth of $50$~MHz.
For more details about the specifications of each module the reader is kindly referred to \cite{manual:NI-5622,chcwbh1201}.

 At the receiver, each antenna is associated with a complete \rf\ chain. For each antenna, the down--converter is used to detect the analogue \rf\ signal from its dedicated antenna. The signal is then sent to the dedicated {16--Bit Digitiser}. The {16--Bit~Digitiser} applies a bandpass filter with a real flat bandwidth equal to  $B_f = (0.4\times f_s)$, where $f_s$ is the sampling rate \cite{manual:NI-5622}. The sampling rate in the experiment is $10$~Ms/s which results in a real flat bandwidth of $4$~MHz. This may result in frequency-selective fading. Nonetheless, equalisation is not required for the detection of \sm\ or \smx\ signals in this experiment because:
\begin{inparaenum}[ i\upshape)]                                                                                                                                                                                                                                                                                                                                                                                                                                                                                                                                                                                                    
 \item there are no multi--tap delays in the experimental setup due to very small distance between the transmit and receiver antennas; and 
 \item \ml\ detection is used to decode the receiver signal for both \sm\ and \smx. The use of \ml\ detection is applied to the complete \sm\ symbol, \ie, the spatial symbol and the signal symbol are decoded jointly.
\end{inparaenum}
 Finally, after synchronisation of the {16--Bit~Digitiser} with the on-board reference clock of the SG--RF, the {16--Bit~Digitiser} writes the received binary files. The simultaneous recording of the two signals coming from Tx1 and Tx2 is facilitated by utilizing multiple processing cores and multiple NI-PXIe modules. The recorded files are then processed according to \dsprx\ in Fig.~\ref{fig:testbed_encoder_decoder_chain}.

 \subsection{Testbed Software} \label{subsec:software}

\begin{figure}
\centering
 \includegraphics[width=8.5cm]{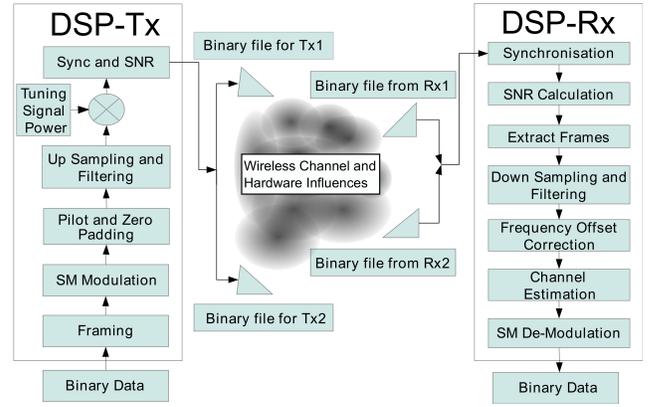}
 \caption{A step-by-step layout of the binary data encoder (DSP--Tx) and decoder (DSP--Rx) processes.}
 \label{fig:testbed_encoder_decoder_chain}
\end{figure}

  Matlab was used to facilitate the digital signal processing required at the transmitter, \dsptx, and at the receiver,\dsprx. \dsptx\ processes the information data and generates binary files that can be transmitted by \pxietx. \dsprx\ process the received data from \pxierx\ and recover the original information data stream. Fig.~\ref{fig:testbed_encoder_decoder_chain} outlines the processing algorithms at \dsptx\ and \dsprx. 

\vspace{0.2cm}

\subsubsection{\textbf{DSP--Tx}}\label{subsubsec:dsptx}

The \dsptx\ process takes the incoming binary information data and performs the following,
\begin{enumerate}[\text{1.}1]
 \item \textbf{Framing}: The incoming data is split into frames consisting of $100$ symbols per frame. 
 \item \textbf{Modulation}: The data in each frame is modulated using \sm\ or \smx:
 \begin{itemize}
  \item \textbf{SM}:  The bit stream is divided into blocks containing ${\log_2\left(N_t\,M\right)}$ bits each, where $M$ is the signal constellation size. The following mapping rule is then used~\cite{mhsay0801}:
\begin{enumerate}
 \item The first $\log_2\left(N_t\right)$ bits determine which transmit antenna is active, \ie, they determine the spatial constellation point of \sm. In this paper, the transmit antenna broadcasting is denoted by $n_t$ with $n_t \in \{1,2,\dots,N_t\}$.
 \item The second $\log_2\left(M\right)$ bits are used to choose a symbol in the signal--constellation diagram. Without loss of generality, \qam\ is considered. The actual complex symbol emitted by the transmit antenna $n_t$ is denoted by $s_t$, with $s_t \in \{s_1,s_2,\dots,s_{M}\} $. 
\end{enumerate}

By following the above steps, the $N_t \times 1 $ dimensional transmit vector is:

\begin{equation}\vspace{0.3cm}
\label{Eq:Mod_vec} 
{\bf{x}}_{n_t ,s_t }  = \left[ {{\bf{0}}_{1 \times \left( {n_t  - 1} \right)} ,s_t ,{\bf{0}}_{1 \times \left( {N_t  - n_t } \right)} } \right]^T , 
\end{equation}

\noindent where $\left[  \cdot  \right]^T$ denotes the transpose operation, and ${\bf{0}}_{p \times q}$ is a $p \times q$ matrix with all--zero entries. Equation \eqref{Eq:Mod_vec} is a representation of the transmission vector for \sm. Since \sm\ activates only one transmit antenna at any transmission instance, it means that only one transmit antenna can broadcast a symbol while all others remain silent. To this extent, the transmit vector is composed of all zeros, except for the single symbol, $s_t$, which is broadcast from antenna $n_t$. In this manner, \sm\ avoids \ici\ and allows  single-stream \ml\ decoding. In addition, \sm\ is energy efficient since only a single \rf\ chain is active, while still providing a multiplexing gain~\cite{ssrhg1201}.

\item \textbf{SMX}:
In this case,  the bit stream is divided into blocks of $N_t\log_2\left(M\right)$ bits, then, according to \cite{f9601}:

\begin{enumerate}
 \item Each $\log_2\left(M\right)$ bits are separately modulated using $M$--\qam\ modulation.
 \item The modulated symbols are then transmitted simultaneously from the $N_t$ transmit antennas.
\end{enumerate}

 \end{itemize}

 \item \textbf{Pilot and Zero Padding}: The \ls\ channel estimation algorithm with local orthogonal pilot sequences is used to estimate the channel \cite{tymj0901}.
 Two pilot signals are added for each frame, one at the start of the frame, and one at the end. Each pilot signal contains ten pilot sequences, where the orthogonal pilot sequence for the $n_t$--th transmit antenna is defined as,
 
 \begin{equation} \vspace{0.3cm}
  \varTheta_{n_t}(\ell) = \exp\left(2\pi j\frac{ n_t \ell}{N_{\varTheta}}\right),
 \end{equation}
 
\noindent where $\varTheta_{n_t}(\ell)$ is the $\ell$--th element of the pilot sequence $\varTheta_{n_t}$ transmitted from antenna $n_t$, $j=\sqrt{-1}$ is the imaginary unit and $N_{\varTheta}$ is the cardinality of the pilot sequence. In this work, the length of each pilot sequence is ${N_{\varTheta}=10}$. 
To avoid \ifi, an all zero sequence  of $50$ zero valued symbols is added to both the start and the end of the frame. Furthermore, a sequence of constant valued symbols is added to enable \fo\ estimation at the receiver. The length of the \fo\ estimation sequence is $1000$ symbols.

\item \textbf{Up Sampling and Filtering}: Up--sampling and matched filtering (pulse shaping) are used to maximise the \snr\ and reduce \isi\ \cite{lm8901}. Each frame is up--sampled with an up-sampling ratio of $4$, and then passed through a \rrc\--\fir\ filter with $40$ taps and a roll--off factor of $0.75$. The large roll--off factor is necessary to ensure that the power is focused in a short time instance to ensure that only a single \rf\ chain is active when using \sm.

\item \textbf{Tuning Signal Power}: The \snr\ is varied by changing the power of the transmitted signal to obtain the \aber. This is done by multiplying each transmission vector with a ``Tuning Signal Power'' factor to obtain the desired transmit power. 
In particular, by changing the amplitude of the ``Data section'' in the transmission vector by using the ``Tuning Signal Power'' factor.

\item \textbf{Synchronisation and SNR}:
Several preamble--autocorrelation based methods for frame synchronisation were tested~\cite{m7201,xj0501,bsio9501}. However, despite the introduction of an interpolation filter at the receiver and due to the channel attenuations, the estimated start of the signal was typically in error by one or two samples. This meant that sample synchronisation could not be achieved consistently, resulting in off-by-one errors. The investigation of synchronisation techniques is outside the scope of this work, but in order to avoid synchronisation via a cable, as is often done in similar experimental systems, the  peak detection technique has been applied which resulted in the desired outcome. We recognise that this technique is suboptimal as it results in power amplifier saturation and potential signal distortions. Nonetheless, a sequence of $20$ symbols with maximum power, separated by $50$ zero valued symbols between each, are added to the start of the transmitted signal. The large power difference between the maximum power peaks and the power of the ``Data section'' symbols is reasonable since the instantaneous channel power may fluctuate by as much as $20$~dB due to fast fading \cite{h9301,mbf0101}. The power difference between the synchronisation section and the remaining sections is set to be larger than the maximum channel variation. In this manner, a successful peak detection is guaranteed. If this is not the case, no peak may be detected at the receiver and all further decoding would be erroneous.

To facilitate \snr\ calculations at the receiver, two sequences of power and no power are added after the synchronisation pulses of the transmitted signal, indicated by ``SNR section'' in Fig.~\ref{fig:Tx_full_vector}. Each sequence contains $5$ blocks of $50000$ symbols and $50000$ zeros. The first sequence is transmitted from the first antenna while the second antenna is off. The second sequence is transmitted from the second antenna while the first antenna is off. 

\end{enumerate}

After the \dsptx\ process completes, the transmit vector symbols are converted to $\mathrm{I}16$ format and are recorded to a binary file. This binary file is then broadcast by \pxietx.

Fig.~\ref{fig:Tx_full_vector} is an absolute value representation of the processed incoming data that is passed to the first transmit antenna~(Tx1) and Fig.~\ref{fig:Tx_single_frame} shows the absolute value representation of each frame. Note that the ``Data section'' is a series of concatenated frames. In Fig.~\ref{fig:Tx_single_frame}, it can be seen that each frame contains $26100$ samples. Therefore, the period of each frame is ${T_{\text{Frame}} = 26100/f_s = 2.6\,\mathrm{ms}}$. This is much less than the coherence time of the channel given that, typically, the coherence time for a stationary indoor environment is approximately $7\,\mathrm{ms}$~\cite[and references therein]{mbf0101}. Hence, the channel estimation at the receiver is valid for the frame duration.

\begin{figure}
 \centering
 \includegraphics[width=8.5cm]{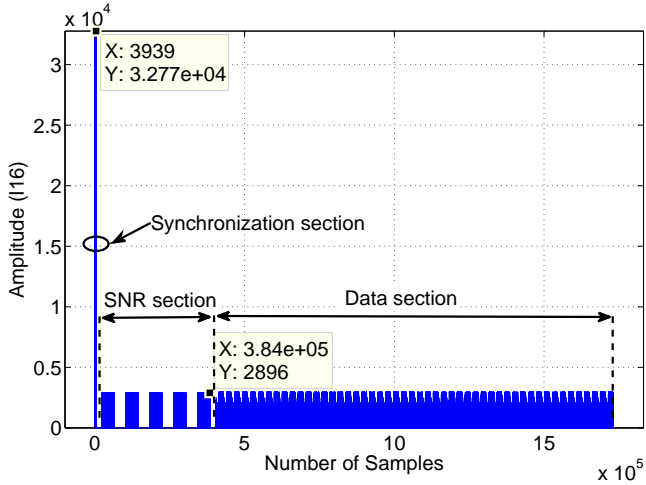}
 \caption[Absolute value representation of the full transmit vector being broadcast on Tx1.]{
 This is the absolute value representation of the transmission vector being sent to $\mathrm{Tx1}$. The synchronisation, \snr\ estimation and Data sections are shown. The value of the peak must equal $2^{15}$ since the $16$bit--Digitiser operates using
an $\mathrm{I}16$ format before tuning the signal power of the data. The highest value in the \snr\ section is the same as the highest value in the information data section, in this example a value of $2896$. The peak value is $2^{15}$. There is approximately a ${21.1\,\mathrm{dB}}$ difference between the peak power in the synchronisation section and the peak power in the \snr\ estimation and data sections. This is apparent when looking at the two data points shown in the figure. }
 \label{fig:Tx_full_vector}
\end{figure}

\begin{figure}
 \centering
 \includegraphics[width=8.5cm]{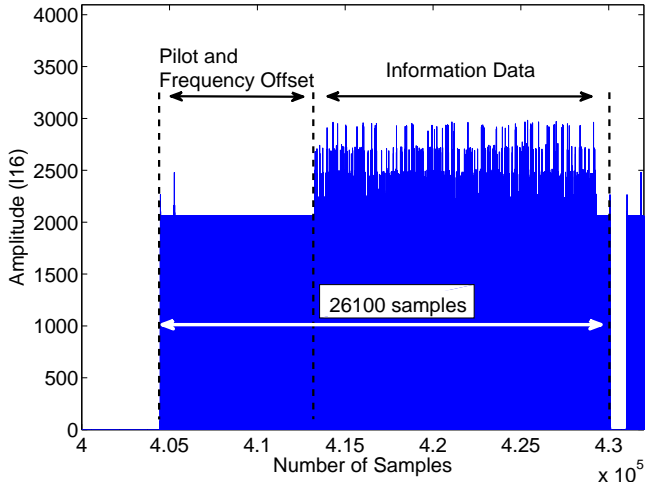}
 \caption[Absolute value representation of a single frame from the transmit vector being broadcast on Tx1.]{ This is the absolute value representation of a single frame from the vector
being transmitted by $\mathrm{Tx1}$ in the $\mathrm{I}16$ data format, which is a signed $15$~bit representation of an integer number.}
 \label{fig:Tx_single_frame}
\end{figure}

\subsubsection{\textbf{DSP--Rx}} \label{subsubsec:dsprx}

The data received by \pxierx\ is processed by \dsprx\ to recover the original data stream. To accomplish this, the following steps are required:

\vspace{0.3cm}

\begin{enumerate}[\text{2.}1]
 
 \item \textbf{Synchronisation}: This is achieved by searching for the peaks with a value above a certain threshold in the received signal. The threshold is set as $70$\% of the highest value in the received vector. This threshold level accounts for the natural voltage variations in the system, \ie, the difference between peak voltage and root-mean-square voltage. If the number of peaks found is less than $20$, then the received vector is discarded from further calculations.
 
 \item \textbf{SNR Calculation}: 
The \snr\ is defined as,

\begin{equation}
\text{SNR}=\frac{\mathrm{E}\left[\left\|\mathbf{Hx}\right\|^2_{\mathrm{F}}\right]}{\sigma_n^2}
\end{equation}

\noindent where $\mathbf{H}$ is the $N_r\times N_t$ channel matrix, $\mathbf{x}$ is the $N_t\times 1$ transmitted vector,  $\mathrm{E}\left[\cdot\right]$ is the expectation operator, and $\left\|\cdot\right\|_{\mathrm{F}}$ is the Forbenius norm.

 Assuming that the noise at the receiver is \awgn, the received signal for the duration of the \snr\ sequence can be written as follows:
 
 \begin{equation}
  \mathbf{y} = \mathbf{h}_{n_t}s_t + \mathbf{n} \label{Eq:Y1}
 \end{equation}
 
\noindent where $\mathbf{y}$ is the $N_r\times 1$ received vector, $\mathbf{h}_{n_t}$ is $n_t$ column of the channel matrix $\mathbf{H}$, $\mathbf{n}$ is the $N_r\times 1$ \awgn\ vector with $\sigma_n^2$ variance and $\mu_n$ mean, and $s_t$ is the transmitted symbol from the $n_t$ antenna. As mentioned in Section~\ref{subsubsec:dsptx}, only a single transmit antennas is active when broadcasting the \snr\ sequence and $s_t$ is either equal to the maximum value in the ``Data section'' $x_{\text{max}}$ or zero, as shown in Fig.~\ref{fig:Tx_full_vector}. Hence, the received signal in \eqref{Eq:Y1} can be re--written as,
\begin{equation}
\mathbf{y} =
\left\{
\begin{array}{ll}
\mathbf{h}_{n_t}x_{\text{max}} + \mathbf{n}, & s_t = x_{\text{max}} \\
\mathbf{n}, & s_t = 0
\end{array}
\right.
\label{Eq:Y2}
\end{equation}
Proceeding from \eqref{Eq:Y2},
\begin{eqnarray}
  \mathrm{E}\left[\left\|\mathbf{Hx}\right\|^2_{\mathrm{F}}\right] &=& \mathrm{E}\left[\left\|\mathbf{y} - \mathbf{n}\right\|^2_{\mathrm{F}}\right]  \label{Eq:Ens}\\
 \sigma_n^2 &=& \mathrm{E}\left[\left\|\mathbf{n}\right\|^2_{\mathrm{F}}\right] - \mathrm{E}\left[\|\mathbf{n}\|\right]^2 
\end{eqnarray}
\noindent where $[\cdot]^{H}$ is the Hermitian operation. As discussed in Section~\ref{subsubsec:dsptx}, each \snr\ sequence contains $50000$ symbols and $50000$ zero valued symbols. Since the noise in the system represents an ergodic process, the ensemble average in \eqref{Eq:Ens} can be replaced with a time average,
 \begin{eqnarray}
\mathrm{E}\left[\left\|\mathbf{Hx}\right\|^2_{\mathrm{F}}\right] &=& \sum_{i=1}^{50000}\left(\left\|\mathbf{y}_i\right\|^2_{\mathrm{F}} - \left\|\mathbf{n}_i\right\|^2_{\mathrm{F}} - 2\mathbf{y}^H_i\mathbf{n}_i\right) \nonumber \\ \\
\sigma_n^2 &=& \sum_{i=1}^{50000} \left\|\mathbf{n}_i\right\|^2_{\mathrm{F}} - \left[\sum_{i=1}^{50000}\|\mathbf{n}_i\|_{\mathrm{F}} \right]^2
 \end{eqnarray}
\noindent where $\mathbf{y}_i$ and $\mathbf{n}_i$ are the $i$--th received vector.
To get a more accurate estimation, the \snr\ is calculated for the $5$ transmitted \snr\ sequences received at both antennas and then averaged again over those measurements.

\item \textbf{Extract Frames}: After finding the start of the transmission and calculating the \snr, \dsprx\ performs a serial to parallel conversion to separate the received frames.

\item \textbf{Down Sampling and Filtering}: To complete the matched filter described in Section~\ref{subsubsec:dsptx}, each frame is down--sampled by a factor of $4$ and passed through an \rrc--\fir\ filter.

\setcounter{figure}{7}
\begin{figure*}[ht]
 \centering
 \includegraphics[width=13.5cm]{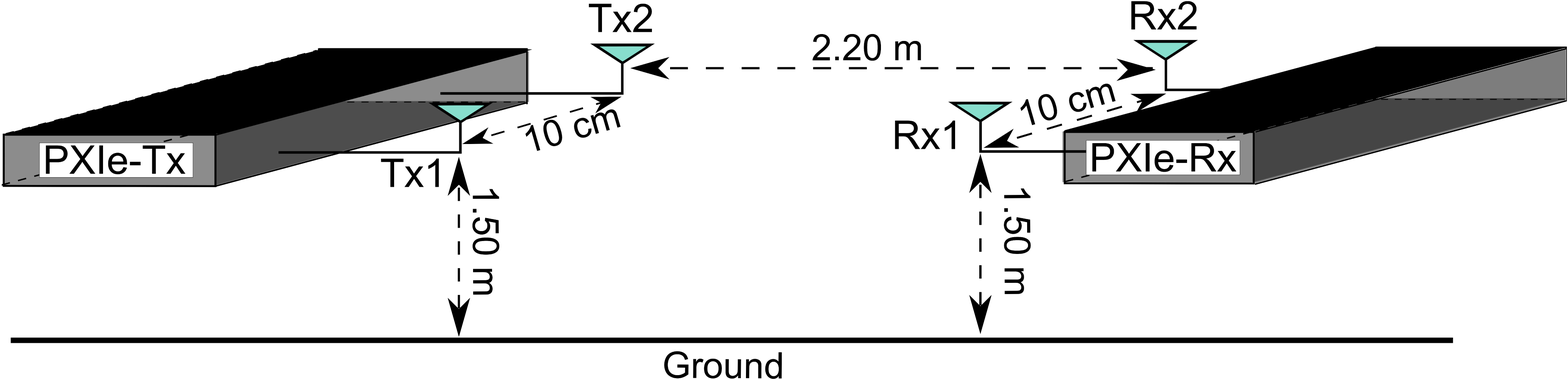}
 \caption[Physical experimental layout.]{Physical experimental layout: A pair of receive and a pair of transmit antennas are set ${2.2\,\mathrm{m}}$ apart from each other with a direct line of sight. Each pair of antennas is ${1.5\,\mathrm{m}}$ from the ground and there is a ${10\,\mathrm{cm}}$ spacing between the antennas in either pair corresponding to $0.77$ times the wavelength at $2.3~\mathrm{GHz}$. All antennas are omnidirectional.}
 \label{fig:testbed_exp_setup}
\end{figure*}

\item \textbf{Frequency Offset (FO) Correction}: The \dsprx\ estimates the \fo\ for each frame by,
\begin{equation}
\label{eq:fo_est}
 \Delta_f = \frac{\angle x_{1000} - \angle x_{1}}{2\pi\times1000}
\end{equation}
\noindent where $\angle x_{1000},\angle x_{1}$ are the angles of the first and the last sample of the \fo\ sequence transmitted by the \dsptx\ where the \fo\ sequence has exactly $1000$ symbols. These angle values are obtained by correcting the radian phase angles in a vector by adding multiples of $\pm 2\pi$ as required. This enables a better estimate of the phase offset. Assuming a linear phase rotation, the frequency offset can be estimated using \eqref{eq:fo_est}. The \fo\ for each frame is then corrected by,
\begin{equation}
\widetilde{y}_i = y_i \times e^{-j2\pi \Delta_f i}
\end{equation}
\noindent where $\tilde{y}_i, y_i$ are the $i$--th element of the corrected and the uncorrected received frame, respectively.

\item \textbf{Channel Estimation}: The channel estimation is done by using the \ls\ channel estimation algorithm proposed in \cite{tymj0901}, where for each frame the channel is estimated by,
\begin{equation}
 \widetilde{\mathbf{H}}_{\text{LS}} = \frac{1}{N_{\varTheta}} \varTheta^H  \mathbf{H}_{r} 
\end{equation}
\noindent where $\mathbf{H}_r$ is the received pilot sequence. To enable a more accurate evaluation of the system, the channel is estimated and averaged over $10$ pilot sequences. Furthermore, two channels are estimated per frame, the first channel estimate is used for the first half of the data symbols in the frame, and the second is used for the second half of the data symbols in the frame.

 \item \textbf{Demodulation}: The \ml\ optimum receiver for \mimo\ systems is used, which can be written as,
\begin{equation} 
\label{eq:MLMIMO}
 \hat{\mathbf{x}}_t^{\left(\text{ML}\right)} =  \mathop {\arg \min }\limits_{\scriptstyle \mathbf{x} \in \mathcal{Q}} \left\{ {\left\| {{\bf{y}} - {\bf{H}}\mathbf{x}} \right\|_{\rm{F}}^2 } \right\} 
\end{equation}
\noindent where $\mathcal{Q}$ contains every possible $\left(N_t\times 1\right)$ transmit vector, and ${\hat  \cdot }$ denotes the estimated transmission vector. However, since only one transmit antenna is active at a time for a \sm\ system, the optimal receiver in \eqref{eq:MLMIMO} can be simplified to,
\begin{equation}
\label{eq:MLSM}
 \left[ {\hat n_t^{\left( {{\rm{ML}}} \right)} ,\hat s_t^{\left( {{\rm{ML}}} \right)} } \right] = \mathop {\arg \min }\limits_{\scriptstyle {n_t} \in \left\{ {1,2, \ldots, N_t } \right\} \hfill \atop
  \scriptstyle s \in \left\{ {s_1 ,s_2 , \ldots, s_M } \right\} \hfill} \left\{ {\sum\limits_{r = 1}^{N_r } {\left| {y_r  - h_{(r, n_t)} s} \right|^2 } } \right\} \\
\end{equation}

\noindent where $y_r$ is the $r$--th entries of ${\bf{y}}$. 

\end{enumerate}

Finally, the recovered binary data along with the estimated \snr\ are used to obtain the \aber\ performance of both \sm\ and \smx.

\setcounter{figure}{6}
\begin{figure}[b!]
 \centering
 \includegraphics[width=8.5cm]{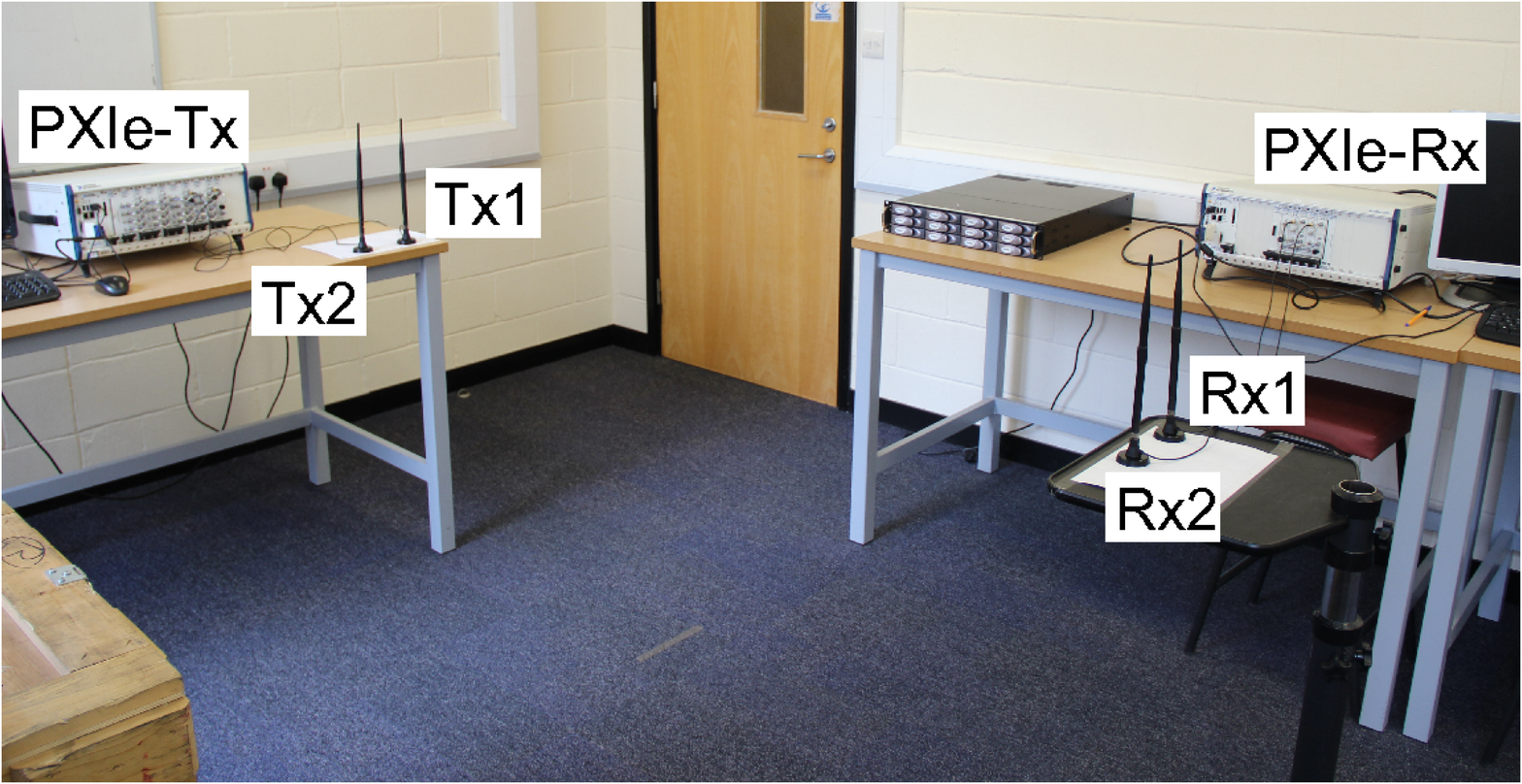}
 \caption{Experimental setup in the laboratory.}
 \label{fig:testbed_exp_lab}
\end{figure}
\setcounter{figure}{8}

\subsection{Propagation Environment (Channel)} \label{subsec:channel_env}

\begin{figure}[ht]
 \centering
 \includegraphics[width=8.5cm]{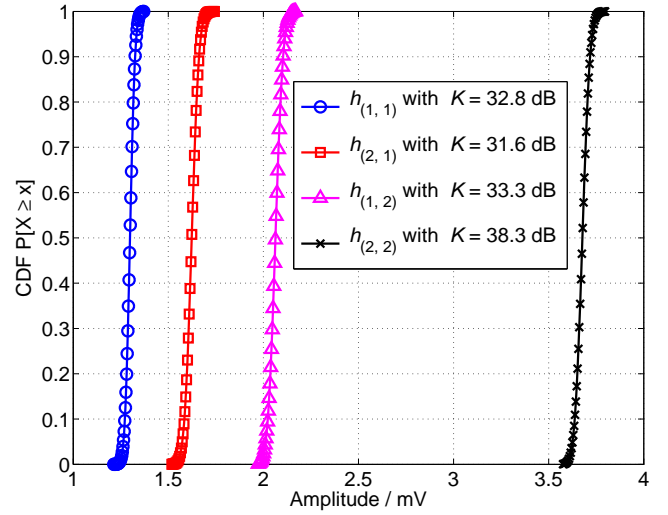}
 \caption[Absolute value representation of a single frame from the transmit vector being broadcast on Tx1.]{\cdf s for each of the fast fading coefficients, $h_{(r, n_t)}$, of the four channels in the experiment. Each is defined by a Rician distribution with a unique $K$-factor. The markers denote the measurement points while the lines denote the best fit approximation. Note that the wireless channel mean values fall in the range of $1.3~\mathrm{mV}$ to $3.6~\mathrm{mV}$.}
 \label{fig:wireless_cdfs}
\end{figure}

The physical layout of the experimental set-up is shown in Fig.~\ref{fig:testbed_exp_lab} and the relative antenna spacing is provided in Fig.~\ref{fig:testbed_exp_setup}. In particular, the two transmit and two receive antennas are identical and are placed directly across from each other. As such, the channel between the transmitter and receiver has a strong \los\ component. Therefore, the channel is assumed to be a Rician fading channel with a large $K$-factor due to the short distance between the transmit and receive antennas, where $K$ is the ratio of the coherent power component, usually the \los, to the non-coherent power components, usually \nlos. The omnidirectional transmit antennas broadcast on a frequency of $2.3$~GHz at $10$~Ms/s.

\begin{figure*}[ht!]
\centering
\begin{minipage}[t]{8.5cm}
  \subfigure[Configuration $(\mathrm{I})$ of the receive \rf\ chains.]{
    \label{fig:coax_cdfs_original}
    \includegraphics[width=8.5cm]{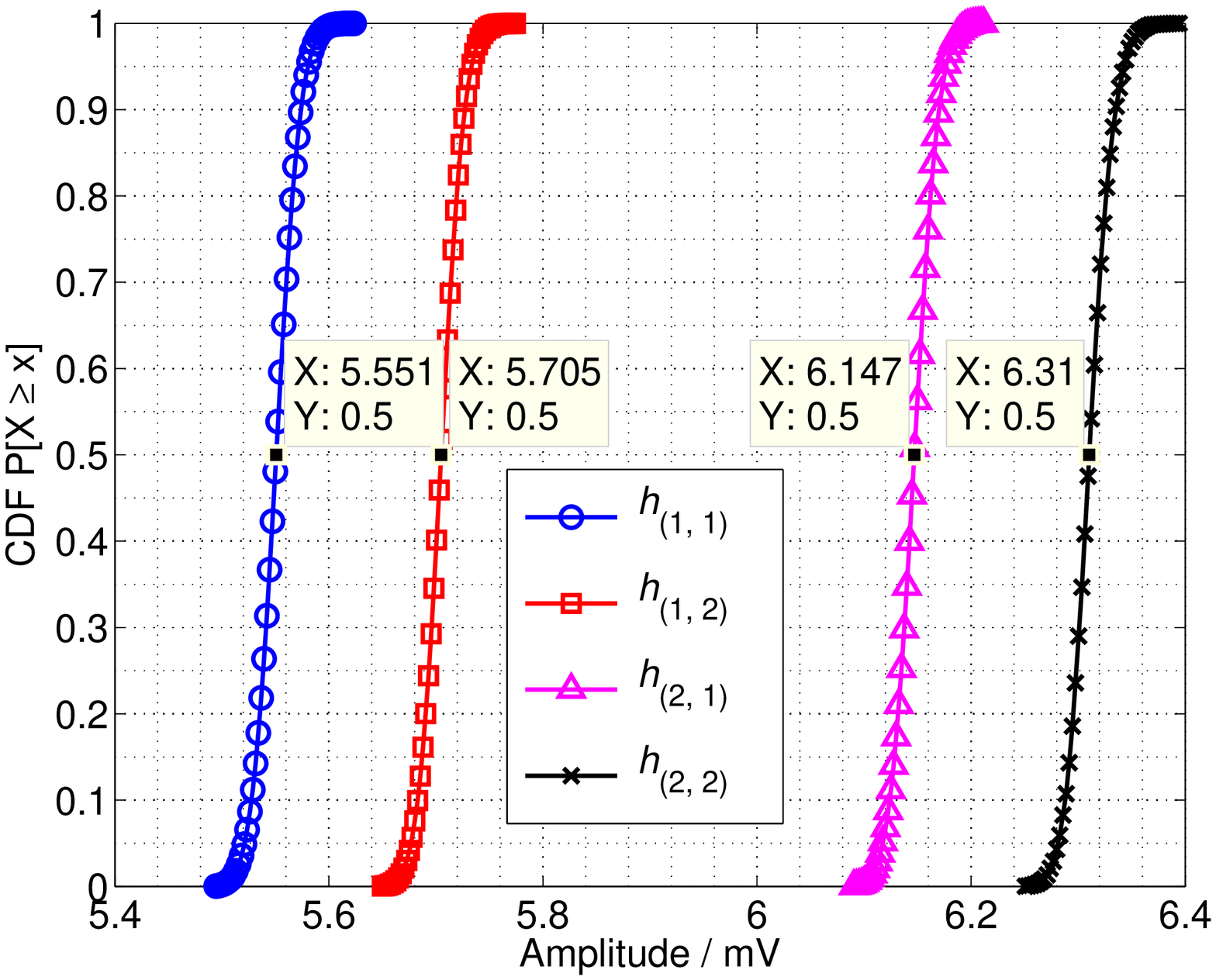} 
  } \\
\end{minipage}
\hfill
\begin{minipage}[t]{8.5cm}
  \subfigure[Configuration $(\mathrm{II})$ of the receive \rf\ chains.]{
    \label{fig:coax_cdfs_swapped}
    \includegraphics[width=8.5cm]{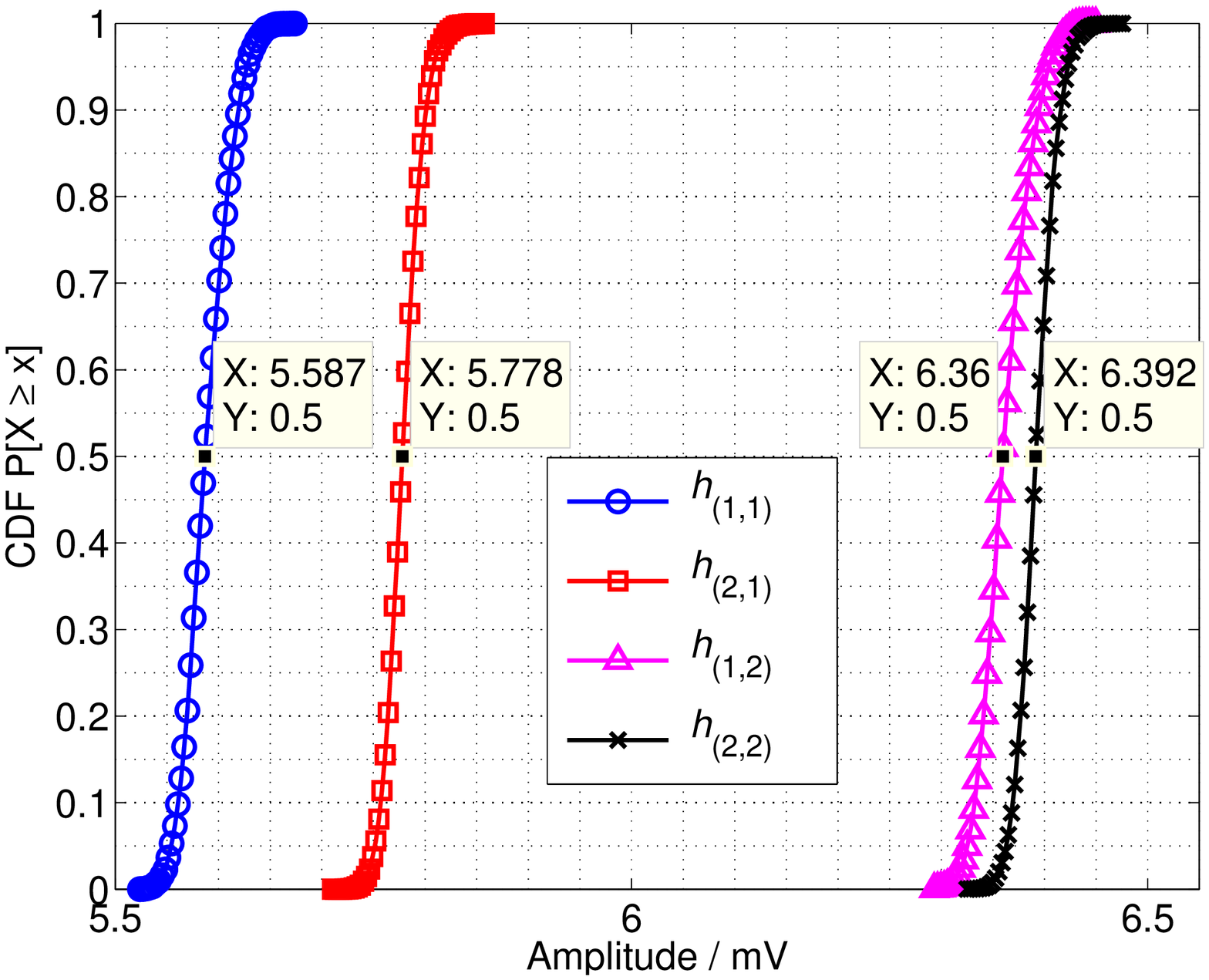}
  }
\end{minipage}
  \caption{\cdf s for each of the fast fading coefficients, $h_{(r, n_t)}$, of the four channels in the experiment. Each is defined by a Rician distribution with a unique $K$-factor. The markers denote the measurement points while the lines denote the best fit approximation. Despite using a coaxial cable with a $10\,\mathrm{dB}$ attenuation to connect the \rf\ chains, each channel exhibits a unique mean.}
  \vspace{-0.1cm}
  \label{fig:coaxial_cdfs}
\end{figure*}

\begin{figure*}[hb]
  \normalsize \hrulefill \vspace*{0pt}

\setcounter{equation}{15}
\begin{equation}
 \label{eq:alpha_factors}
\alpha_{(1, 1)} = 0 \,\mathrm{dB}, \quad \alpha_{(2, 1)} = 0.25 \,\mathrm{dB}, \quad \alpha_{(1, 2)} = 0.88 \,\mathrm{dB},
\quad \alpha_{(2, 2)} = 1.1 \,\mathrm{dB}, 
\end{equation}

\begin{equation}
 \label{eq:alpha_factors_swapped}
 \alpha_{(1, 1)} = 0 \,\mathrm{dB}, \quad \alpha_{(2, 1)} = 0.29 \,\mathrm{dB}, \quad \alpha_{(1, 2)} = 1.13 \,\mathrm{dB},
\quad \alpha_{(2, 2)} = 1.17 \,\mathrm{dB}. 
\end{equation}

\setcounter{equation}{14}
\end{figure*}

Channel measurements were collected to verify that the channel environment followed a Rician distribution. To achieve this, the transmitter broadcasts pulses at $10$~Ms/s on a carrier frequency of $2.3$~GHz at $4$~dBm peak power. Each pulse includes a frequency offset estimation section and a total of $10^5$ pulse samples are collected. A best fit approximation is then calculated for the collected data. In particular, a maximum likelihood estimation is fitted to the collected data. A Chi-squared goodness-of-fit test is then performed to ascertain that the distribution resulting from the maximum likelihood estimation fits at least $95\%$ of the data. The empirical \cdf\ for each link is presented in Fig.~\ref{fig:wireless_cdfs}. The results show that the channel does follow a Rician distribution with a $K$ factor that ranges between $31-38$ dB. The different $K$-factors on the links between the transmit and receive antennas may be explained by the room geometry, the antenna  positioning and the overall propagation environment. However, note that each of the \cdf s has a different mean, which will be discussed in the next section.

\section{Equipment Constraints} \label{sec:testbed_constraints}

Fig. \ref{fig:testbed_exp_setup} shows the physical layout of the experiment. Note that the $10$~cm inter--antenna separation used here is sufficient to guarantee very low, if any, spatial correlation when broadcasting at $2.3$~GHz with a $2.2$~m separation between the transmitter and receiver~\cite{qohd0901}. 

 The physical environment through which the signal passes, starting from the SG--RF at the transmitter, until it reaches the $16$--Bit Digitisers at the receiver, suffers from connector losses, differences in the \rf\ chains, different phase responses, attenuations and similar. To study and model the effects of the hardware imperfections on the signal power:
 \begin{itemize}
  \item An \rf\ coaxial cable with a $10$~dB attenuation is connected between each transmit and receive antenna. 
  \item A pulse is transmitted at $10$~Ms/s on a carrier frequency of $2.3$~GHz at $-10$~dBm peak power. Each pulse includes a frequency offset estimation section and a total of $10^5$ pulse samples were collected. 
  \item The \cdf\ for each of the fading coefficients is calculated and is shown in Fig.~\ref{fig:coaxial_cdfs}.
 \end{itemize}

 In an ideal environment, the means of the \cdf s in Fig.~\ref{fig:coaxial_cdfs} should be equal. However, imperfections in the hardware result in different means for each transmit to receiver antenna pair, as can be seen in Fig.~\ref{fig:coaxial_cdfs}. The differences between the channels can be modelled as a \pim\ between the various link pairs in the channel matrix $\mathbf{H}$. Therefore, the channel coefficients are redefined as,
 \begin{equation}
  \label{eq:redefined_channel_coef}
  h^{\text{PI}}_{\left(r,n_t\right)} = \sqrt{\alpha_{\left(r,n_t\right)}} \times h_{\left(r,n_t\right)}
 \end{equation}
\noindent where $\alpha_{\left(r,n_t\right)}$ is the channel attenuation coefficient from receive antenna $r$ to transmit antenna $n_t$. 
 
 To locate the source of the discrepancy between the different channel attenuations, \ie, determine if the NI modules or the NI chassis is the source, the \rf\ chains at the receiver were swapped around and the channels were estimated in configuration $(\mathrm{I})$ and configuration $(\mathrm{II})$.
 To clarify, configuration $(\mathrm{I})$ represents the default modular set-up of the testbed while configuration $(\mathrm{II})$ refers to swapping the front-end modules around the transmit chassis.
 Fig.~\ref{fig:coax_cdfs_original} shows the channel \cdf\ for each transmit to receive antenna pair in configuration $(\mathrm{I})$ while Fig.~\ref{fig:coax_cdfs_swapped} shows the channel \cdf\ for each transmit to receive antenna pair in configuration $(\mathrm{II})$.
 By considering the means of the \cdf s in Figs.~\ref{fig:coax_cdfs_original} and~\ref{fig:coax_cdfs_swapped} and taking $h_{(1,1)}$ as a base, the various channel attenuations that result when the receiver is in configuration $\mathrm{I}$ or in configuration $\mathrm{II}$ are given in \eqref{eq:alpha_factors} and \eqref{eq:alpha_factors_swapped} respectively.
 Comparing Fig.~\ref{fig:coax_cdfs_original} and Fig.~\ref{fig:coax_cdfs_swapped}, as well as the attenuations in \eqref{eq:alpha_factors} to those in \eqref{eq:alpha_factors_swapped}, shows that they are very similar. Indeed, swapping of the \rf\ chains has a minimal impact on the estimated mean of each channel attenuation. Thus, it can be assumed that the NI modules that compose the receive \rf\ chains are the source of the hardware imperfections, and consequently lead to the differences in the means of the estimated \cdf s. To account the hardware imperfections, the channel attenuation coefficients in \eqref{eq:alpha_factors} and \eqref{eq:alpha_factors_swapped} are taken into consideration in the derivation of the analytical model in Section~\ref{sec:testbed_ana}. The accuracy of the derived analytical bound using the channel attenuation coefficients in \eqref{eq:alpha_factors} and \eqref{eq:alpha_factors_swapped} is demonstrated in Section~\ref{sec:testbed_results} where it is compared with empirical results.

\setcounter{equation}{17}

\section{Analytical Modeling}
\label{sec:testbed_ana}

\vspace{0.1cm} 

An analytical model for the \aber\ performance of the experimental system is developed by considering the system model presented in Section~\ref{sec:testbed_sys_model} and the system constraints in Section~\ref{sec:testbed_constraints}. The performance of \sm\ and \smx\ over a single link in a noise-limited scenario is characterised by 

\begin{equation}
\label{eq:ber_sm_sums}
\mathrm{ABER} \leq  \frac{1}{2^m}\sum_{\mathbf{x}_t}\sum_{\mathbf{x}}\frac{N\left({\bf{x}}_{t},{\bf{x}}\right)}{m}\mathrm{E}_{\mathbf{H}}\left\lbrace\underset{\text{error}}{\Pr} \right\rbrace ,
\end{equation}

\noindent where $N\left({\bf{x}}_{t},{\bf{x}}\right)$ is the number of bits in error between the transmitted vector ${\bf{x}}_{t}$ and ${\bf{x}}$, $\mathrm{E}_{\mathbf{H}}\{\cdot\}$ is the expectation across the channel $\mathbf{H}$, and $\underset{\text{error}}{\Pr}$ is the conditional \pep\ of deciding on ${\bf{x}}$ given that ${\bf{x}}_{t}$ is transmitted~\cite{rh1003}, 

\begin{eqnarray}
 \Pr_{\text{error}} &=&  \Pr\left(\left\| \mathbf{{y}} - \mathbf{{H}}{\bf{x}}_{t}\right\|^2_{\mathrm{F}} > \left\| \mathbf{{y}} - \mathbf{{H}}{\bf{x}}\right\|^2_{\mathrm{F}}   \bigg | \mathbf{H} \right)  \nonumber \\
 &=&  Q\left(\sqrt{\gamma_\mathrm{ex}\left\|\mathbf{{{H}}}\left({\bf{x}}_{t}-{\bf{x}}\right)\right\|^2_{\mathrm{F}} }\right) \label{eq:pep} 
\end{eqnarray} 

\noindent where $\gamma_\mathrm{ex} = \frac{E_m}{2N_0}$ is half of the \snr\ between the transmitter and receiver, and ${Q(\omega) = \frac{1}{\sqrt{2\pi}} \int_\omega^\infty \exp\left(-\frac{t^2}{2}\right) \mathrm{dt} }$ is the $Q$-function. As Fig.~\ref{fig:testbed_exp_setup} indicates, the transmit and receive antennas in the experiment experience a very strong \los\ environment. Accordingly, the channel between each transmit to receive antenna pair is characterised by Rician fading. A generic Rician channel is defined as

\begin{equation}
\label{eq:rician_fade}
 h_{(r, n_t)} = \sqrt{\frac{K}{1+K}}  + \sqrt{\frac{1}{1+K}} \widetilde h_{(r, n_t)} ,
\end{equation}

\noindent where ${\widetilde h_{(r, n_t)}\sim\mathcal{CN}(0, 1)}$ is a complex normal, circular symmetric random variable with zero mean and unit variance. $n_t\in\{1,2\}$ is the index of the transmit antenna and $r\in\{1,2\}$ is the index of the receive antenna. 

To account for the hardware imperfections that result from the power imbalances, the fast fading channel coefficients are redefined according to~\eqref{eq:redefined_channel_coef},~\eqref{eq:alpha_factors}, and~\eqref{eq:rician_fade}.
Section~\ref{sec:testbed_results} validates the derived analytical bound by comparing it to experimental and simulation results. 

\section{Complexity Analysis} \label{sec:complexity}
\vspace{0.1cm} 

 The computational complexity of the \ml\ detector for \sm\ (\sm--\ml) is compared to that of the \ml\ detector for \smx\ (\smx--\ml). The complexity is computed as the number of real multiplicative operations $(\times,\div)$ needed by each algorithm. The detailed derivation of each expression is considered in~\cite{ysrmh01} and references therein.

\begin{itemize}
 
 \item \smx--\ml: The computational complexity of the \smx--\ml\ receiver outlined in \eqref{eq:MLMIMO} is equal to,
 
   \begin{equation}
      \label{eq:CompSMX} 
      \mathcal{C}_{\text{SMX--ML}} = 4\left(N_t+1\right)N_r2^m ,
   \end{equation}
   
\noindent  where $m$ is the spectral efficiency of the system. Note that $\left(\left|\mathbf{y}-\mathbf{Hx}\right|^2\right)$ in \eqref{eq:MLMIMO} requires $\left(N_t+1\right)$ complex multiplications.

 \item  \sm--\ml: The computational complexity of the \sm--\ml\ receiver outlined in \eqref{eq:MLSM} is equal to, 
 
  \begin{equation}
     \label{eq:CompSM}
     \mathcal{C}_{\text{SM--ML}} = 8N_r2^m  ,
  \end{equation} 

  \noindent where the \ml\ detector searches through the entire transmit and receive search spaces. Note that evaluating the Euclidean distance $\left({\left| {y_r  - h_{(r, n_t)} s_t} \right|^2 }\right)$ requires 2 complex multiplications, where each complex multiplication requires 4 real multiplications. 

\end{itemize}

 Considering \eqref{eq:CompSMX} and \eqref{eq:CompSM}, for the same spectral efficiency, the reduction in complexity of the \sm--\ml\ receiver relative to that of the \smx--\ml\ receiver is given by,
 
 \begin{equation}
  \mathcal{C}_{\text{rel}} = 100 \times \left( 1 - \frac{2}{N_t+1} \right) \label{eq:RelComp}
 \end{equation}

\begin{figure}
    \centering
      \includegraphics[width=9.2 cm]{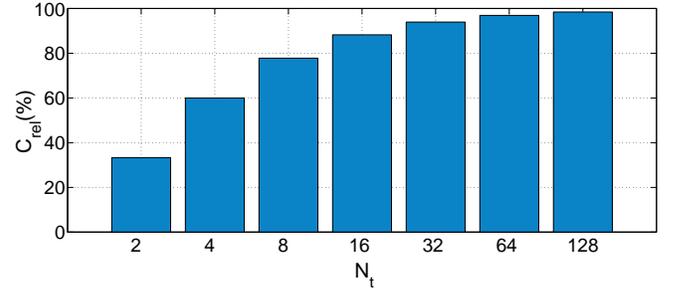}
      \caption{Relative receiver complexity reduction of the SM--ML receiver versus the SMX--ML receiver.}
    \label{fig3:CompChart}
\end{figure}

On the one hand, as can be seen in \eqref{eq:CompSM}, the complexity of the \sm\ receiver does not depend on the number of transmit antennas, and it is equal to the complexity of \simo\ systems. On the other hand, the complexity of \smx\ increases linearly with the number of transmit antennas. Therefore, as the number of transmit antennas increases, the relative gain of the \sm\ receiver increases. This can be seen in Fig. \ref{fig3:CompChart} where the relative complexity for ${N_t \in \{2,4,8,\ldots,128\}}$ is shown for both systems. In fact, Fig.~\ref{fig3:CompChart} shows that for ${N_t=4}$, \sm\ offers a $60\%$ reduction in complexity, while a $98\%$ reduction in complexity can be seen for ${N_t=128}$. The theoretical, simulation and empirical results for \sm\ and \smx\ are now discussed.

\section{Experimental Results and Numerical Analysis} \label{sec:testbed_results}

\subsection{Measurement Campaign} \label{subse:measurement_camp}
A stream of $10^5$ information bits is sent per transmission to obtain the experimental results. Two transmit antennas are available and \bpsk\ is used for the signal constellation. As mentioned in Section~\ref{subsubsec:testbed_physical_Rx}, the real flat bandwidth is $4$~MHz. The information data is put in $50$ frames with $2000$~bit each, as shown in the ``Data section'' of Fig.~\ref{fig:Tx_full_vector}. The channel is estimated at the beginning and the end of every frame, resulting in $100$ channel estimations per transmission vector. The experiment is repeated $1000$ times for every \snr\ point. In addition, analytical and simulation \aber\ curves are shown for \sm\ in a Rician environment with and without the \pim s given in \eqref{eq:alpha_factors}. 

\subsection{Results} \label{subsec:results}
\begin{figure}[ht!]
 \centering
 \includegraphics[width=8.8cm]{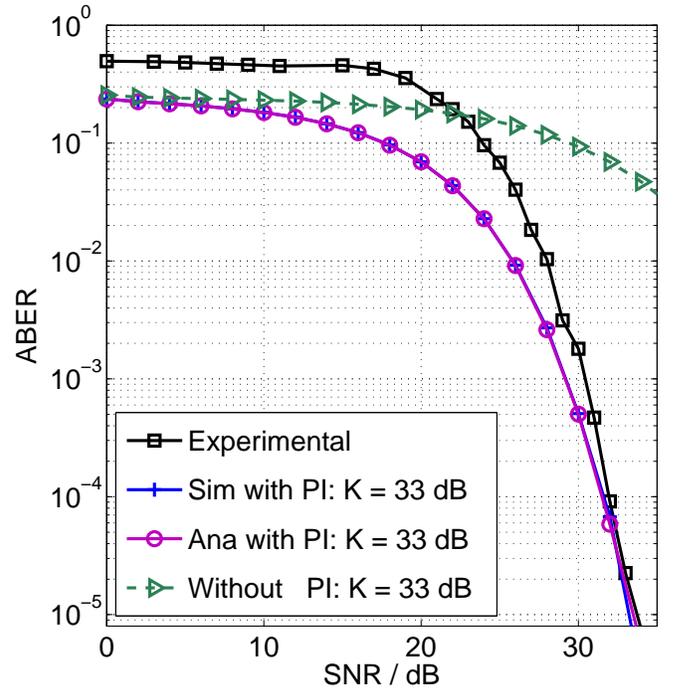}\vspace{0.1cm}
 \caption[ABER for SM in the experimental set-up.]{\aber\ for \sm\ in an experimental set-up with $2$ transmit antennas, $2$ receive antennas and a spectral efficiency of ${2\, \mathrm{bits/s/Hz}}$. The \snr\ is set as measured on $h_{(1,1)}$ with ${\alpha_{(1,1)} = 0\, \mathrm{dB}}$. The solid black line with square markers denotes the experimental results. The green diamond markers denote simulation results with no power imbalance (\pim) between the links while the green dashed line is the analytical prediction. The remaining curves denote the simulation $(\mathrm{Sim})$ and analytical $(\mathrm{Ana})$ results.}
   \vspace{-0.4cm}
 \label{fig:SM_BER_testbed}
\end{figure}

The simulation, analytical and experimental results for the \aber\ performance of \sm\ in a \los\ channel are illustrated in  Fig.~\ref{fig:SM_BER_testbed}. In particular, the experimental results approximate the performance of the simulation results with \pim s and both the simulation, and experimental results, are closely approximated by the derived upper bound at a low \aber. 

This result serves to validate theoretical work done in the field where the presented \snr\ along the x--axis is equivalent to the \snr\ on $h_{(1,1)}$. The large error between the experimental, simulation and analytical curves at high \aber\ can be attributed to a number of factors including incorrect frequency offset estimation, timing recovery errors, synchronisation problems, poor channel estimation and decoding. Notably, incorrect frequency offset estimation can result in a systematic error contributing significantly to the $30$\% error seen at low \snr s in the figure. As the \snr\ increases, however, frequency offset estimation, timing recovery and channel estimation improve, leading to a lower \aber\ as shown in Fig.~\ref{fig:SM_BER_testbed}. Differences between the measured and simulated \aber\ curves can be attributed to channel imperfections such as channel correlations, mutual coupling and interference signals from the surrounding environment. Quantifying these imperfections is deemed important and requires channel modelling and interference measurement. However, addressing these effects is beyond the scope of this work and will be subject of future works.

\sm\ performs best in a rich scattering environment where the channel between each transmit and receive antenna is unique. In particular, the larger the Euclidean distance between two received vectors is, the better the \aber\ performance of \sm\ becomes. Conversely, the more similar the channels are, the worse the \aber\ of \sm\ is. However, the channel uniqueness can be the result of the scattering environment or \pim s caused by hardware tolerances. The analytical and simulation results presented in Fig.~\ref{fig:SM_BER_testbed} show the poor performance of \sm\ in a Rician environment with no \pim\ between the various transmitter to receiver links. Fig.~\ref{fig:SM_BER_testbed} also shows the analytical and simulation \aber\ for \sm\ when \pim\ are introduced. Indeed, the \aber\ of \sm\ improves significantly when these \pim s are introduced as each channel becomes more separable. This increases the Euclidean distance and improves performance. 

If the channels between each transmit antenna to each receive antenna are similar, then the \aber\ performance of \sm\ degrades. This is seen when looking at the \sm\ system without \pim s, illustrated by the dashed green line with triangular markers in Fig.~\ref{fig:SM_BER_testbed}. In fact, the \aber\ of \sm\ can be approximated by separating the error that originates from the estimation of the spatial constellation symbol and the error that originates from the estimation of the signal constellation symbol \cite{dh1101}. Therefore, depending on the environment, the main contributor to the overall \aber\ of a \sm\ system will be the erroneous detection of the spatial or signal constellation.

When \pim s are introduced, the Euclidean distance between the channel signatures increases. This decreases the error contribution of the spatial component of \sm. Hence, when the \snr\ is sufficiently high to have near perfect channel estimation, the error of the system is bound by the error from the signal component of \sm. 
This separation can only be shown when iterative detection is used, which is proven to be sub-optimal \cite{jgs0801}. In addition, work in \cite{rh1101_ssk_rician} shows that the error when only the spatial constellation of \sm\ is used for data transmission gets worse for an increasing $K$ factor in a Rician environment. This is the opposite to conventional modulation techniques since a larger $K$ factor for \sm\ means a smaller Euclidean distance between the spatial constellation points which results in an increased \aber\ performance. Indeed, it is the Euclidean distance between the different channels that determines the error in the spatial constellation detection. However, since \ml--optimal detection is used at the receiver, separating the error from the spatial and signal symbols is strictly not permitted. Please note that the \pim s between the links are always obtained relative to the channel with the greatest attenuation, \ie, the values of the \pim\ factors in \eqref{eq:alpha_factors} and \eqref{eq:alpha_factors_swapped} are always positive.

\begin{figure}[ht]
 \centering
 \includegraphics[width=8.8cm]{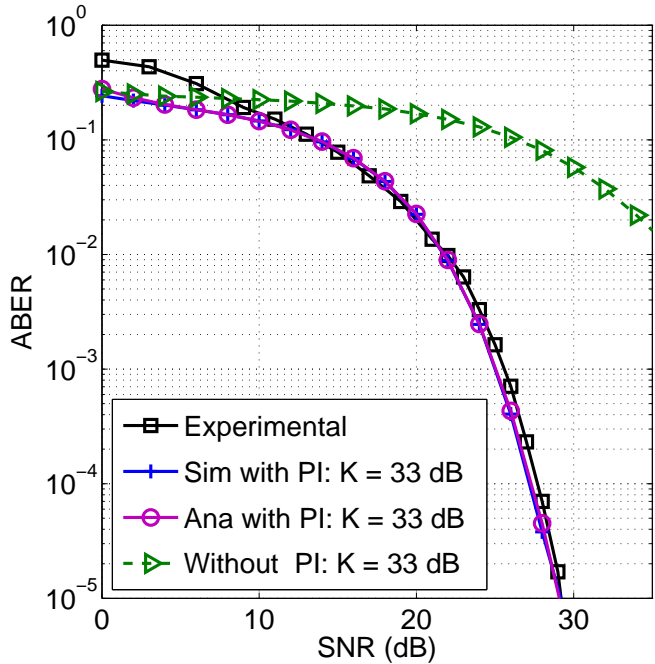}
 \caption[ABER for SMX in the experimental set-up.]{\aber\ for \smx\ in an experimental set-up with $2$ transmit antennas, $2$ receive antennas and a spectral efficiency of ${2\, \mathrm{bits/s/Hz}}$. The \snr\ is set as measured on $h_{(1,1)}$ with ${\alpha_{(1,1)} = 0\, \mathrm{dB}}$. The solid black line with square markers denotes the experimental results. The green diamond markers denote simulation results with no power imbalance (\pim) between the links while the green dashed line is the analytical prediction. The remaining curves denote the simulation $(\mathrm{Sim})$ and analytical $(\mathrm{Ana})$ results.}
 \label{fig:MIMO_BER_testbed}
\end{figure}

Furthermore, power imbalances between the transmitting antennas are shown to offer improved performance in terms of the \aber\ when only the spatial constellation of \sm\ is used, \ie, when \ssk\ is the underlying modulation technique. In particular, an optimised power allocation for a various number of transmit antennas is addressed in \cite{rh1006}, where the authors show that there is optimal power allocation between the transmitting antennas which can serve to increases the Euclidean distance between the channel signatures and improve the \aber\ performance of \sm. Indeed, \sm\ has also been successfully applied to an \awgn\ optical wireless channel where it is shown that \pim s greatly improve the \aber\ performance \cite{fhrm1101}.

\begin{figure}[!h]
 \centering
 \includegraphics[width=8.8cm]{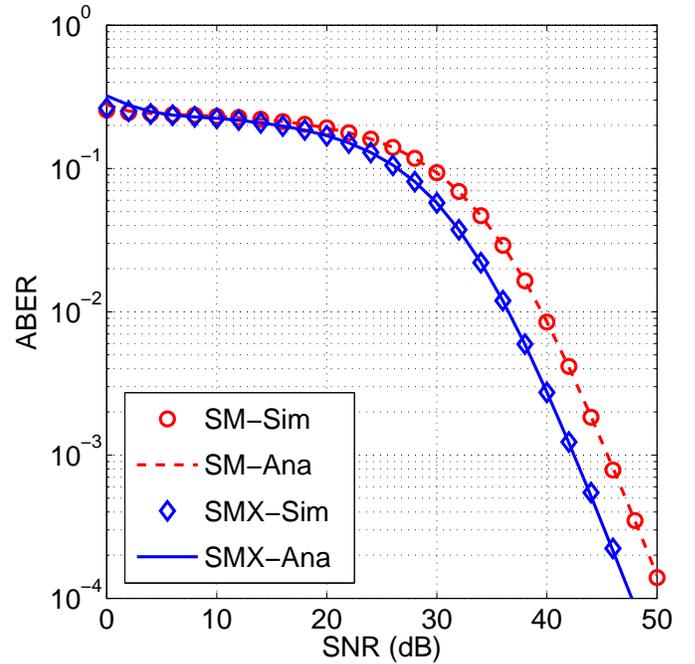}
 \caption[ABER for SM and SMX in theory.]{\aber\ for \sm\ and \smx\ in a Rician fading channel where ${K = 33~\mathrm{dB}}$ with $2$ transmit antennas, $2$ receive antennas, a spectral efficiency of ${2\, \mathrm{bits/s/Hz}}$ and no \pim s between the channels. The simulation $(\mathrm{Sim})$ are denoted by markers while the analytical $(\mathrm{Ana})$ results are denoted by the lines.}
 \label{fig:BER_SM_MIMO_theory}
\end{figure}

The simulation, analytical and experimental results for the \aber\ performance of \smx\ in a \los\ channel are illustrated in Fig.~\ref{fig:MIMO_BER_testbed}. In particular, the experimental results closely follow the performance of the simulation results with \pim s and both the simulation, and experimental results, are closely approximated by the derived upper bound at low \aber\ when the hardware imperfections are taken into account. This result serves to validate theoretical work done in the field. The results in Fig.~\ref{fig:MIMO_BER_testbed} demonstrate that the \smx\ system, like the \sm\ system, also benefits from the \pim s in the hardware. The \smx\ system exhibits approximately a $3~\mathrm{dB}$ coding gain when compared to \sm\ at an \aber\ of $10^{-4}$. This coding gain can also be seen at an \aber\ of $10^{-3}$ in Fig.~\ref{fig:BER_SM_MIMO_theory}, where the simulation and analytical results for the \aber\ performance of \sm\ and \smx\ are shown when there are no \pim s between the links. 

\begin{figure}[!h]
 \centering
 \includegraphics[width=8.8cm]{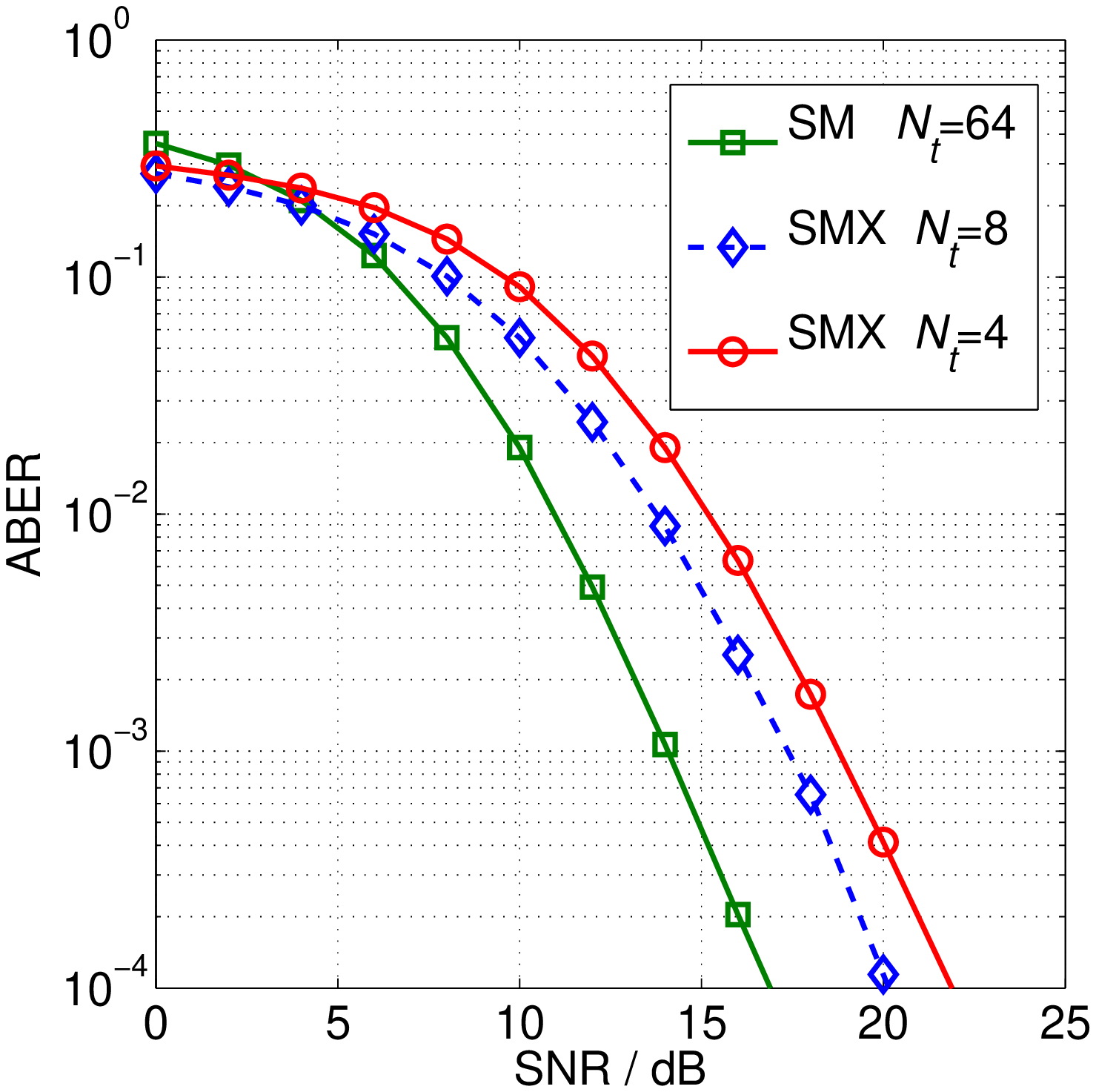}
 \caption[ABER for SM and SMX in theory.]{Simulation results for the \aber\ for \sm\ and \smx\ in a Rayleigh fading environment with a spectral efficiency of ${8\, \mathrm{bits/s/Hz}}$ and no \pim s between the channels.}
 \label{fig:BER_large_Nt}
\end{figure}

The coding gain of \smx\ relative to \sm\ is expected when there are few transmit antennas. The Euclidean distance between the transmit vectors, and therefore the variance in~\eqref{eq:pep}, in \smx\ is larger than in \sm. However, the aim of this paper is to show that empirical results validate the simulation and analytical work done in the field, which can be seen in both Fig.~\ref{fig:SM_BER_testbed} and Fig.~\ref{fig:MIMO_BER_testbed}. Unfortunately, due to the limited number of transmitter and receiver \rf\ chains available, there are no experimental results for systems with a larger number of transmit or receive antennas where \sm\ is shown to perform better than \smx. These empirical results will be the focus of future research. Nonetheless, the accuracy of the theoretical and simulation results of \smx\ and \sm\ with a large number of transmit and receive antennas can be extrapolated from the presented results.

Fig.~\ref{fig:BER_large_Nt} compares the \aber\ between \sm\ (solid lines) and \smx\ (dashed lines) in a system with a large number of transmit antennas. Each system operates in a Rayleigh fading environment with a spectral efficiency of ${8\,\mathrm{bits/s/Hz}}$ and four receive antennas. The results demonstrate the coding gains available to a \sm\ system as compared to \smx\ when a large number of transmit antennas are available. In particular, \sm\ with $N_t=64$ offers a coding gain of up to $4$~dB with respect to \smx\ with $N_t=8$ and a coding gain of $6$~dB with respect to \smx\ with $N_t=4$. These performance gains stem from the greater Euclidean distance between the transmit vectors for \sm. It is important to note that although \sm\ is simulated as having $64$ transmit antennas available, it requires only a single \rf\ chains, while \smx\ requires $8$ \rf\ chains for the $8$ transmit antennas. Furthermore, to achieve the \aber\ performance illustrated in Fig.~\ref{fig:BER_large_Nt}, \sm\ requires $64$ unique channels. In this work, a unique channel is assumed to be available only with the addition of a single transmit antenna. However, work in \cite{ysmh1001,bapp1101} and others, look at creating multiple channel signatures without the need for a large number of physical transmit antennas while maintaining a similar \aber\ performance to the traditional \sm\ scheme. 

This work demonstrates that the hardware tolerances of practical communication systems are beneficial for the \aber\ performance of both \sm\ and \smx. This behaviour along with the requirement for a single \rf\ chain, make \sm\ a viable candidate for future wireless networks.

\section{Summary and Conclusion}
\label{sec:testbed_conclusion}

In this work, the \aber\ performance of \sm\ and \smx\ has been validated experimentally for the first time. In particular, the encoding and decoding processes were presented. The experimental testbed, equipment and channel conditions were then described in detail and 
the \aber\ of \sm\ and \smx\ were obtained in a practical testbed environment. In addition, the experimental results were compared to both simulation and analytical approaches. As a result, it has been shown that a Rician channel with different channel attenuations closely described the behaviour of \sm\ and \smx\ in the physical environment. Furthermore, it was demonstrated that the different channel attenuations resulted from various hardware imperfections at the transmitter and receiver \rf\ chains. In fact, the induced power imbalances resulted in significant coding gains for the practical systems relative to the theoretical predictions without such power imbalances. To this extent, \sm\ and \smx\ performed as expected relative to the theoretical work when the power imbalances were introduced in the analytical model. This result validated the \sm\ principle. The performance gains exhibited by \sm\ in the practical implementation make \sm\ a viable candidate for future wireless networks and particularly for systems with a large number of transmit antennas available. 

It is worth noting that the presented work may be extended in a number of different ways that would broaden its applicability. Empirical results that demonstrate the performance of \sm\ and \smx\ with a large number of transmit and receive antennas remain to be obtained. In light of the above results, the \aber\ performance of \sm\ and \smx\ is expected to follow the theoretical models, but these results are essential to validate the \aber\ performance for both \sm\ and \smx\ systems. In addition, channel imperfections such as channel correlations and mutual antenna coupling along with the impact of interfering signals from neighbouring transmitters on the same frequency, should be analysed. Furthermore, obtaining empirical results for the capacity and energy efficiency of \sm\ are of great interest for future research, particularly since \sm\ is projected to have large energy efficiency gains when compared to other traditional \mimo\ schemes  since it requires only a single RF chain. As a consequence, the quiescent power and circuit power can be kept at low levels. Acquiring the hardware which would enable the accurate measurement of these aspects is key. Lastly, the implementation of the \sm\ detection algorithm on a DSP or an FPGA brings with it a number of optimisation challenges such as the use of multi-threading, pipeling, fixed point computations and others. The deployment of \sm\ on an FPGA or a DSP has yet to be demonstrated.

It has been shown that \sm\ is a simple, low cost, \mimo\ technique, which has now demonstrated excellent performance in a \los\ wireless channel. Therefore, this work shows that \sm\ is a promising practical approach to obtaining the enhanced performance of spatial multiplexing without introducing high processor complexity and high power consumption that would occur when using other spatial multiplexing approaches. The aim now is to investigate the performance of \sm\ in a range of experimental channel conditions and further study its potential.

\section*{Acknowledgement}

We gratefully acknowledge partial support by the University of Edinburgh Initiating Knowledge Transfer Fund~(IKTF), EPSRC Fellowship~(EP/K008757/1), RCUK~(EP/G042713/1,UK-China Science Bridges "R\&D on (B)4G Wireless Mobile Communications"), the European Union~(PITNGA2010264759, "GREENET"), and the Key Laboratory of Cognitive Radio and Information Processing~(Guilin University of Electronic Technology), Ministry of Education, China~(Grant No.: 2013KF01).

\bibliographystyle{IEEEtran}
\bibliography{IEEEabrv,general,cwc}

\end{document}